# Van der Waals Engineering of Valley Polarization in $WSe_2$ via Kagome $V_2O_3$ Monolayer Heterostructure through Magnetic Proximity Effect

Fazle Subhan[1], Kiduk Kim[1], Jaeyoung Kim[1], Guangzhao Qin[*,2], Luqman Ali[3], and Joonkyung Jang[*,1]

1. *Department of Nanofusion Technology, Pusan National University, Busan 46241, Republic of Korea*
2. *State Key Laboratory of Advanced Design and Manufacturing Technology for Vehicle, College of Mechanical and Vehicle Engineering, Hunan University, Changsha 410082, P. R. China*
3. *Department of Physics, University of Virginia, Charlottesville, VA, 22911, USA*

## ABSTRACT

To achieve a robust valley splitting in transition metal dichalcogenides (TMDs) at zero magnetic field, is a challenging and elusive for valleytronic and spintronic devices. Therefore, in the current study, using the first principles calculations, we introduce an oxide-driven valleytronic platform by employing a two-dimensional ferromagnetic oxide ($V_2O_3$) as a magnetic proximity partner for WSe2. We found that the intrinsic ferromagnetism of $V_2O_3$ induces a prominent and spontaneous valley splitting of ~ 10.41 meV in $WSe_2$. This pronounced valley polarization mainly originates from the strong interfacial exchange coupling interaction and charge redistribution mediated by V 3d electrons, coupled with the intrinsic spin-orbit coupling of $WSe_2$. Interestingly, this value significantly greater than the previously reported value for $CrI_3/WSe_2$ which corresponds to an effective magnetic field of ~ 10 T. Besides, we also have a high Curie temperature of 500 K, and an out-of-plane magnetic anisotropy energy (MAE) of 0.31 meV, indicating that this oxide based heterostructure can also be used near-room temperature operation. Importantly, under the external electric field with a step of ±0.1 eV/Å, still the ferromagnetism preserves and enhancement in Curie temperature and MAE make this oxide-based heterostructure more valuable for the next generation valleytronic devices. Therefore, these findings establish a new paradigm for realizing tunable, robust, and magnetic field free valleytronic and spintronic devices for this oxide based heterostructure. Moreover, this concept can be generalized to other correlated oxide based TMDs systems, providing a versatile strategy for next generation quantum and functional materials.

**Correspondance :** jkjang@pusan.ac.kr (J. J), gzqin@hnu.edu.cn (G. Q)

# I. INTRODUCTION

The family of two-dimensional (2D) materials has enormously grown which unlock the new paradigms and accelerating the investigation of potential phenomena with rich physics. The emergence in the two-dimensional magnetic materials, such as $CrI_3$[1], $VI_3$[2], $Cr_2Ge_2Te_6$[3] and $Fe_3GeTe_2$[4] up to the low-dimensionality, also trigger the researchers to explore new insights using the van der Waal's (vdW) heterostructure. These possible new 2D materials and their vdW heterostructure exponentially increase the huge expansion of different materials in condensed matter physics for last few years. Interestingly, the band alignment in the these vdW heterostructure causes many ideal platforms and enable new developments such as manipulation of spintronic,[5] excitonic,[6] topological[7] and the superconducting[8] phenomenon. Apart from the individual materials in 2D monolayers, the stacking in different forms can also help to achieve improved functions.[9] The valley polarization can be controlled in $YI_2$, $LaI_2$ and $GdBr_2$ in their 2D van der Walls bilayers.[10] Furthermore, some of the well-established theories, Berezinskii-Kosterlitz-Thouless (BKT) transition of the XY model,[11,12] Mermin-Wagner theorem[13] and Ising transition[14] also provide a way to thoroughly study the magnetic vdW heterostructure. Interestingly, the modern semiconductor heterostructure plays a vital role in industries and technologies due to their prominent properties in electronics, opto-electronics and valleytronics devices.[15] Although, there is no enthusiastic efforts has been made in past for exfoliating and potentially using them as a substrate in the individual monolayer except for the theoretical efforts.

In graphene like 2D materials, owing to the quantum Hall effects[16] and the breaking of time reversal symmetry, the two inequivalent corners of Brillion zone (±K), gives another prominent phenomenon called valley pseudospins. Since valleys emerge at the corners of the hexagonal Brillion zone as two degenerated states, they offer significant potential for device applications.[17,18] Experimentally, circularly polarized optical pumping is employed to selective excite valley pseudospins, while a high magnetic field induces Zeeman splitting which are leading to the breaking of time reversal symmetries. Furthermore, the optical Stark effect provides an alternative way to lift the valley degeneracy by inducing an effective large magnetic field, even if we have no external magnetic field. Interestingly, the presence of valley degeneracy in the momentum space of transition metal dichalcogenides (TMDs) leads to a variety of an emerging physical phenomena, such as valleytronics,[19–21] valley Hall effects[22] and the valley dependent photoluminescence.[23] In 2D materials, the time reversal symmetry enforces the energetic degeneracy and coupling the two inequivalent valleys, therefore its breaking can profoundly alter the electronic properties.[24] Interestingly, Mak et al. experimentally demonstrated the valley Hall effect in $MoS_2$ monolayer, where carriers are originating from the different valleys with a propagation in the opposite transverse directions.[25] Mostly, due to the strong spin-orbit coupling (SOC) effect in the Group VI TMD's and breaking of time reversal symmetry the valleytronics, and Rashba effect occurred. Besides different approaches such as stacking effect,[26] twisting and gate voltage,[27] external magnetic field,[28] creating vacancies,[29] and doping some magnetic ions,[30] strain,[31] and layer engineering approaches[32,33] in the vdW heterostructure can also be used for valley splitting. Similarly, doping in $Fe_2Se_2O$ monolayer and perpendicular Neel vector in colinear antiferromagnetic materials such as $Co_2S_2$ can lead them to multifunctional spintronic and valleytronic devices.[34,35]Further, the layer polarized spin Hall effect (LP-SHE) in bilayers can also make a 2D material as an ideal candidate for spintronic applications.[36] Among, all these approaches, van der Waal's effect for inducing valley splitting[37] and the magnetic proximity effect in the vdW heterostructure are very prominent.[27,38]

Moreover, the 2D van der Waal's materials can not only widen and enhances the valley studies but their engineering with a TMD 2D material also broaden the spin-dependent charge hopping across the interface of vdW heterostructure by the excitation helicity dependent photoluminescence intensity.[39] Interestingly, some researchers also confirmed the dependency of valley splitting over the interlayer distance,[26,40] while others have reported on the interlayer twisting.[41] Besides, Zhang et *al.* and Imran et *al.* investigated the stacking dependent approach for lifting the valleys in heterostructure.[26,42] These studies clearly prove that valley manipulation in TMDs, can be easily achieved from a ferromagnetic substrate which causes a proximity effect. This proximity effect is a straightforward and a promising approach for valley polarization.

The heterostructure is a prominent resurgence of interest for 2D magnetic materials so that to fabricate a heterostructure of a TMD and ferromagnetic monolayer and overcome the challenges for applications in electronics, valleytronics and spintronics. However, the oxide materials or more specifically vanadium oxide, still needed to be explored both experimentally and theoretically. Therefore, in our current study, we explored the first principles calculations for valley splitting in $WSe_2$ monolayer on another 2D magnetic oxide substrate called $V_2O_3$ monolayer and additionally explored the magnetic proximity effect. Because both electrons and holes don't have an equal space extremum at $K^+$ and $K^-$ but instead they have degenerate momentum space extremum at $K^+$ and $K^-$. Thus, the central issue in the current study is enhanced valley splitting and the giant magnetic anisotropy with the perpendicular electric field.

## II. NUMERICAL METHOD

In the present study we use the density functional theory calculations with projected augmented wave (PAW) under the Vienna ab initio simulation package (VASP),[43,44] with the generalized gradient approximation (GGA) adopted in Perdew-Burke-Ernzerhof (PBE).[45] For the plane wave kinetic energy cut-off of 500 eV was used. The Brillouin zone has been sampled by the Monkharst-Pack method [46] within atomically generated k-mesh of 11x11x1. The structural optimization of the $V_2O_3/WSe_2$ heterostructure was carried out using the DFT-D3 van der Waal's correction, and the optimized geometry was subsequently used for all electronic, magnetic, and structural analysis.[47] The convergence criterion for energy was set to $10^{-5}$ eV, while the atomic positions were optimized until the Hellmann-Feynman force on each atom becomes smaller than 0.01 eV/Å. To describe the strong correlation effects of the localized V 3d electrons in $V_2O_3$, the DFT+U method was employed. The on-site Coulomb interaction was treated using the Dudarev approach, where only the effective Hubbard parameter $U_{eff} = U-J$ is considered. In the present calculations, $U_{eff}$ = 3.68 eV was applied to the V 3d orbitals.[48,49] For magnetic anisotropy energy (MAE), a non-collinear total energy calculation with spin orbit coupling (SOC) was adopted for determining the magnetization direction. The energy criterion was set up to $10^{-8}$ and a denser k-mesh of 21x21x1 was used with and without external electric field. The convergence of magnetic anisotropy energy was carefully checked along the two magnetic axes [1 0 0] and [0 0 1] which are called parallel (in-plane) and perpendicular (out-of-plane). The energy difference between these two directions ($E_{in} - E_{out}$) gives us magnetic crystalline anisotropy energy (MAE). To avoid the artificial interactions between the images more than 20 Å vacuum along the z-axis were applied. Further, to obtain the Curie temperature ($T_c$), we calculated the temperature dependent magnetization curve based on the Metropolis Monte Carlo (MC) simulations using the VAMPIRE software package.

## III. RESULTS

### A. Crystal Structure of monolayers

In the current study, as depicted in Figure 1 (a - b), in $V_2O_3$ monolayer (ML), where vanadium (V) atoms are surrounded by three O atoms in a hexagonal form, also these O atoms form a honeycomb Kagome lattice structure. The lattice constant of the $V_2O_3$ ML is 6.193 Å where the slight reduction in the lattice constant of the $V_2O_3$ ML after structural relaxation, is because of the difference in their atomic radii. Interestingly, the V-to-V atom distance is 3.576 and similarly for O-O atoms is 3.098 Å. Further, the relaxed crystal structure of $V_2O_3$ ML belongs to a space group of P6/MMM (D6H-1) with # 191. We already report about the formation energy and binding energy in our previous report about the $V_2O_3$ ML.[49] Besides the formation and binding energies, we also checked the stability of $V_2O_3$ ML using the phonon and thermodynamic stability in our previous study.[49] Similarly, the hexagonal transition metal dichalcogenide $WSe_2$ is also reported. As $WSe_2$ exists in two phases called trigonal (1T) and hexagonal (2H) under the normal conditions of temperature and pressure.[50] The 2H phase is energetically more stable; therefore, we consider the 2H phase in our calculations. After fully structural optimization, our relaxed lattice constant (3.32 Å) for $WSe_2$ monolayer agrees well with the experimentally reported values.[51] The top and side view for the $WSe_2$ monolayer is given in the Electronic Supplementary Information Figure-ESI-1 (a)-(b) respectively.

### B. $V_2O_3/WSe_2$ heterostructure

Now, to construct the van der Waals (vdW) $V_2O_3/WSe_2$ heterostructure, the optimized lattice constants of the isolated $V_2O_3$ and $WSe_2$ monolayers (MLs) were 6.193 Å and 3.321 Å, respectively. To reduce the lattice mismatch between the two layers, a 1×1×1 unit cell of $V_2O_3$ and a 2×2×1 supercell of $WSe_2$ (6.642 Å) were employed to build the heterostructure. The lattice mismatch between the two monolayers can be calculated by using the relation below

$$\delta = (a^H - a^M)/a^M \times 100\ \% \quad (1)$$

Where $a^H$ and $a^M$ denote the lattice constants of the $V_2O_3/WSe_2$ heterostructure and $V_2O_3$ (or $WSe_2$) MLs respectively. Based on this relation, the lattice mismatch between $V_2O_3$ and the 2×2x1 $WSe_2$ supercell is approximately 6.87%. To form a commensurate interface, a common in-plane lattice constant of 6.45 Å was adopted. Consequently, the $WSe_2$ layer undergoes a compressive strain of −2.89% as its lattice constant decreases from 6.642 Å to 6.45 Å, while the $V_2O_3$ monolayer experiences a tensile strain of about 4.15%, expanding from 6.193 Å to 6.45 Å. The presence of such strain can significantly influence the interlayer coupling and orbital hybridization at the interface, thereby influencing the electronic and structural properties of the $V_2O_3/WSe_2$ heterostructure. Particularly, previous studies shows that compressive lead to a notable reduction in band gap of $WSe_2$ reaching values of approximately 1.457 eV at around 3% strain.[52] Furthermore, strain-induced modifications in crystal field symmetry and orbital overlap can also change the spin-orbit coupling (SOC) at the interface and thus lead to an affective change in the magnetic anisotropy energy (MAE), while the applied strain can also indirectly alter the thermal stability of the magnetic ordering in $V_2O_3/WSe_2$ heterostructure. Nevertheless, the dominant mechanism responsible for the observed valley splitting is attributed to the magnetic proximity effect rather than strain alone.[53,54]

To determine the most stable configuration of $V_2O_3/WSe_2$ heterostructure, we consider several possible stackings as shown in Electronic Supplementary Information Figure ESI-2. The relative stability of each configuration can be easily evaluated from the total energy calculations. To compare the ground state energies of different stackings, the energy of S1 configuration was taken as the reference (0 meV). The calculated energy differences for S2, S3 and S4 configurations are 24.48, 76.26 and 77.34 meV respectively, relative to S1. In contrast to the $WSe_2$ monolayer, both in $V_2O_3$ monolayer and $V_2O_3/WSe_2$ heterostructure, we included the van der Waal's (vdW) interactions. We also calculated the interlayer distance between V-W and V-Se for different stackings before and after relaxation. The calculated values for the interlayer distance are shown in the Electronic Supplementary Information Figure ESI-3 (a)–(b). Consequently, we discovered that O on W top, also known as S1 stacking, is the most advantageous of the other stackings in terms of both energies and interlayer distance.

Therefore, S1 configuration will be further investigated for electronic, magnetic, and interfacial properties of $V_2O_3/WSe_2$ heterostructure. Besides, we also explore that the bond lengths between O-V (α) and W-Se (β) before (after) relaxation are respectively 1.88 (1.867 Å) and 2.54 (2.537 Å) which shows no prominent change, while the interlayer distance $d_0$ before (after) the structural optimization is 2.256 (3.138 Å) in the S1 stacking as shown in Figure 1 (a)–(b) for top and side view respectively.

We also calculated the interlayer binding energy using the relation below

$$E_b = E_{tot} - E_{WSe_2} - E_{V2O3} \qquad (2)$$

where $E_b$ is the binding energy and $E_{tot}$ total energy of the $V_2O_3/WSe_2$ heterostructure and $E_{WSe2}$ and $E_{V2O3}$ are the energies of each monolayer respectively. The total energy of heterostructure, $WSe_2$ and $V_2O_3$ respectively is $E_{tot}$ = -128, $E_{WSe2}$ = - 86.4 and $E_{V2O3}$ = - 41.1 eV, which results in a binding energy of $E_b$ -133.58 meV. As we have a total of 17 atoms in unit cell of the $V_2O_3/WSe_2$ heterostructure, while the binding energy per atom becomes -7.86 meV which is in accordance with some DFT calculated reports for different vdW heterostructures.[26,42] Finally, we apply a perpendicular external electric field in a step of 0.1 V/ Å to the $V_2O_3/WSe_2$ vdW heterostructure. The electric field is applied in two opposite directions called out of plane or positive (From $WSe_2$ towards $V_2O_3$ ML) and in-plane or negative (From $V_2O_3$ towards $WSe_2$ ML) respectively. In literature, a high impact of the electric field over the electronic and magnetic properties of the 2D $V_2O_3/WSe_2$ heterostructure has been reported.[42,55] The calculated results for interlayer distance are given in Table 1.

C. **Electronic Properties**

The central issue in our current calculations is to investigate the valley splitting due to the magnetic proximity effect by the $V_2O_3$ ML. Therefore, first we need to explore the electronic properties of $V_2O_3/WSe_2$ vdW heterostructure using the GGA+U technique. For this purpose, we investigate the spin and orbital resolved band structure as shown in Figure 1(c)–(d) respectively. Interestingly, Figure 1 (c)–(d) depicts that we have a direct band gap of 0.03 eV for $V_2O_3/WSe_2$ vdW heterostructure with a Dirac point on the high symmetry K point. Where both the CBM and VBM are located at K point. Further, from the spin resolved projected band structure of the $V_2O_3/WSe_2$ vdW heterostructure, we analyzed that only the majority spin of V atom is responsible for the Dirac point, while all the other atoms have negligible contribution as shown in Figure 1 (c). Further, orbital resolved band analysis also confirms that the Dirac point mainly originated from the majority spin of V-$d_{yz}$ orbital as shown in Figure 1 (d).

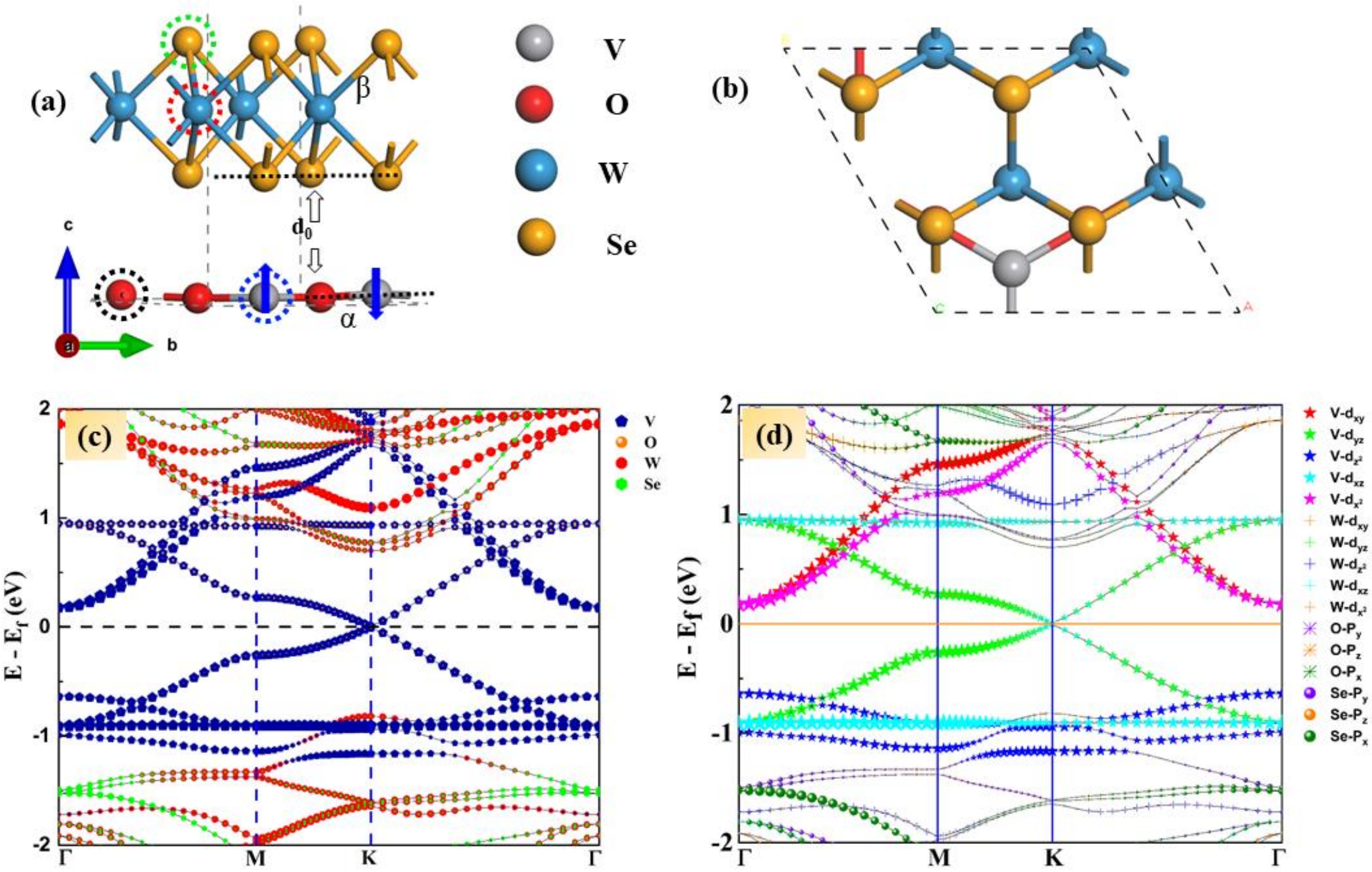

**Figure 1:** Structural illustration for $V_2O_3/WSe_2$ vdW heterostructure on (a) side and (b) top view, respectively. The gray, red, sky blue, and orange balls respectively show the V, O, W and Se atoms. (c-d) GGA+U electronic band structure of $V_2O_3/WSe_2$ vdW heterostructure for (c) majority spin and (d) orbital resolved in majority spin alignment.

We also calculated the minority spin alignment both in the spin and orbital resolved for $V_2O_3/WSe_2$ vdW heterostructure as well as the atomic resolved projected band structure using the GGA+U. These results are respectively depicted in Electronic Supplementary Information Figures ES I–4 and 5.

Next, we explore the effect of external electric fields over the electronic band structure in positive and negative directions. Interestingly, the direct band gap nature preserves with increase in the electric field irrespective of the electric field direction. Our results are given in Table 1 for positive and negative directions.

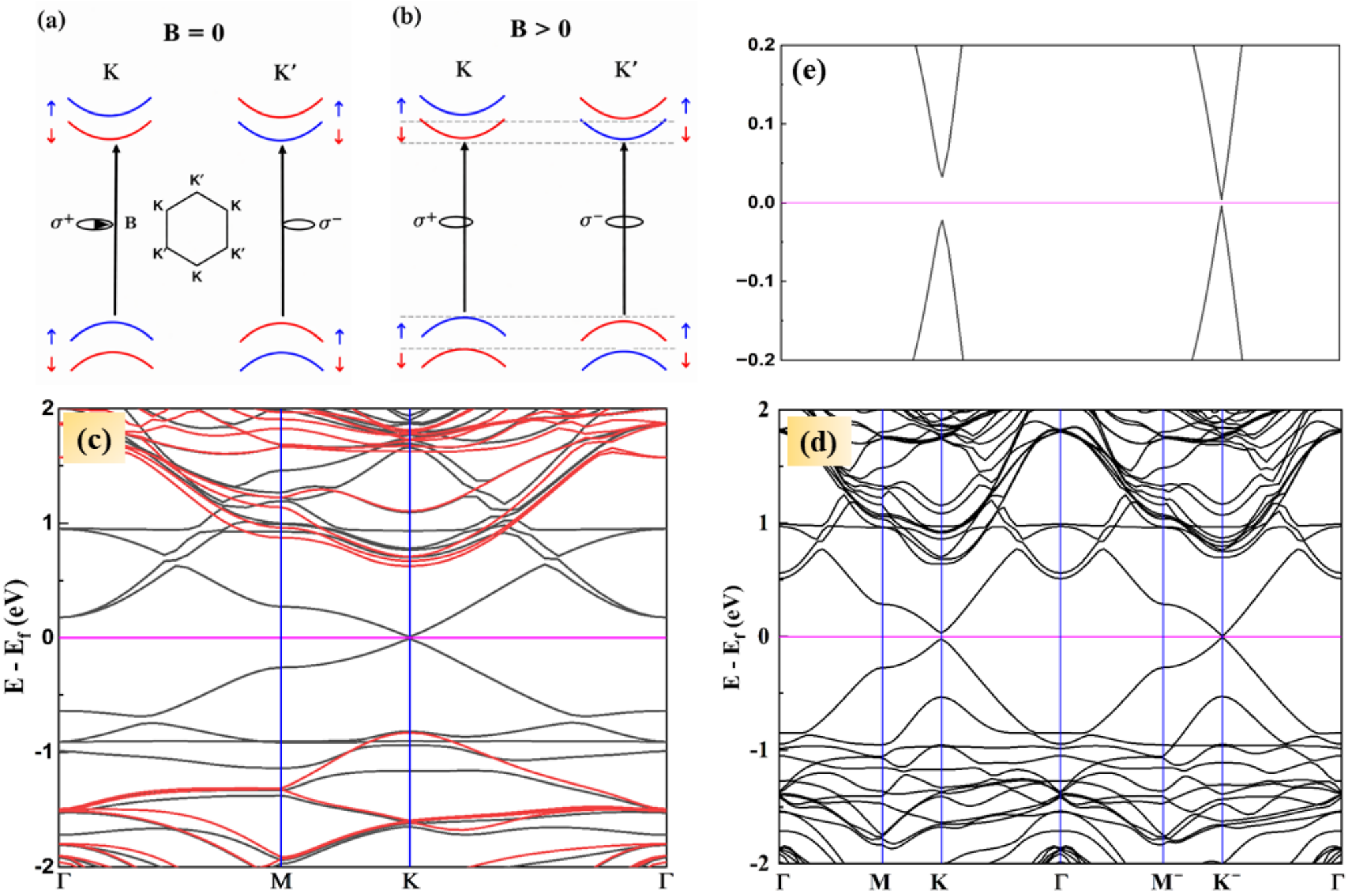

**Figure 2:** A schematic diagram of the band structures for valley dependent optical selection rule in (a) $WSe_2$ ML and (b) for $V_2O_3/WSe_2$ vdW heterostructure at $K^+$ and $K^-$ due to magnetic proximity effect, where **B** shows the induced magnetic field due to $V_2O_3$ ML. (c, d) The PBE+U and SOC calculated band structure of $V_2O_3/WSe_2$ vdW heterostructure at $K^+$ and $K^-$. (e) Magnified view of the band structure highlighting the band gaps at K and $K^-$ valleys respectively.

As we know that among different 2D materials, some of the transition metal dichalcogenides (TMDs) including $WS_2$, $WSe_2$, $MoSe_2$ and $MoS_2$ are some important 2D materials for investigating the valleytronics properties due to a peculiar characteristic of broken and preserved inversion symmetry in the odd and even number of layers respectively.[15,28,56,57] In the present work, the main issue is to explore the valley polarization. Since we discuss earlier that during the heterostructure formation, we found an appreciable compressive strain of -2.89 % in $WSe_2$ monolayer, therefore a question arises as to whether the current valley splitting originates from the strain effect or from the magnetic proximity effect induced due to the $V_2O_3$ monolayer. To address this issue, first we calculate the effect of biaxial compressive strain (-2.89 %) on the valley splitting of isolated $WSe_2$ monolayer. The results are shown in the Electronic Supplementary Information Figure ESI-6. Although, this strain only reduces the electronic band gap with a change from direct to indirect band gap nature. Furthermore, we found no lifting of the valley degeneracy. Thus, we conclude that magnetic proximity effect induced by the magnetic $V_2O_3$ monolayer is responsible for valley splitting as discussed in detail below.

Next, we need to achieve this goal of valley splitting, which is mainly due to the magnetic proximity effect induced by the $V_2O_3$ substrate. For this purpose, first it is essential to understand the valley splitting in terms of twofold valley degree of freedom as shown in Figure 2(a)– (b). As shown in Figure 2(a), in the presence of broken inversion symmetry and strong spin-orbit coupling, the interband optical transitions follow valley dependent optical selection rules, allowing $\sigma^+$ ($\sigma^-$) circularly polarized optical transitions at $K^+$ and $K^-$.[58,59] Consequently, the two valleys exhibit equivalent optical transition energies for circularly polarized light with an equivalent energy $\left(E_{\sigma^+} = E_{\sigma^-}\right)$. Furthermore, at the conduction band maximum (CBM), the bands at K (C1, C2) and K' (C'1, C'2) are very close to each other with an energy difference of 34 meV. Similarly, for VBM these band states at K (V1, V2) and K′ (V′1, V′2) are slightly far away with an energy difference of 467 meV as given in the Electronic Supplementary Information Figure ESI-7.

Thus, this phenomenon results in equivalent energy for the circularly polarized light $\left(E_{\sigma^+} = E_{\sigma^-}\right)$ as shown in Figure 2 (a). Further, the SOC calculated band structure of $WSe_2$ monolayer is also presented in Electronic Supplementary Information Figure ESI-7, where both the valleys at $K^+$ and $K^-$ with circularly polarized light having an equivalent energy $\left(E_{\sigma^+} = E_{\sigma^-}\right)$. Here the energy difference between the two valleys at $K^+$ and $K^-$ ($\Delta_1\uparrow$ - $\Delta_2\uparrow$) is zero. While in case of an external magnetic field **B**, not only the inversion symmetry broken but the time reversal symmetry also breaks down and thus their energies will no longer remains equivalent $\left(E_{\sigma^+} \neq E_{\sigma^-}\right)$ as shown in Figure 2 (b) and thus the spins and valley can be distinguished. Similarly, in case of $V_2O_3/WSe_2$ heterostructure, the magnetic proximity effect produced by the ferromagnetic $V_2O_3$ monolayer induces valleys behave as an external magnetic field **B**. Therefore, the energy corresponding to the $\sigma^+$ circularly polarized light is 0.0104 eV at $K^+$ while for $\sigma^-$ is -0.00015 eV at $K^-$ as shown in Figure 2 (d). Furthermore, Figure 2 (d) reveal that band-edge states associated with the valley degree of freedom are located at K and $K^-$ with a finite energy difference at $K^+$ and $K^-$ ($\Delta_1\uparrow$- $\Delta_2\uparrow$) is 10.41 meV which confirms the valley splitting due to the broken inversion and time reversal symmetries. Interestingly, from these calculations we confirm that the substrate ferromagnetic material ($V_2O_3$ ML) induces valley splitting like the magnetic proximity effect by the external magnetic field **B**.

To estimate the required external magnetic field **B**, we can use the k.p model[60], where the interaction energy is given by the following equation

$$H_B = g_0 \mu_B \boldsymbol{B}.\boldsymbol{S} + \mu_B \boldsymbol{B}.\boldsymbol{L} \qquad (3)$$

Where **B** shows the external magnetic field, $g_0$ is the Landé factor having a value of 2, $\mu_B$ is the Bohr magneton and $\mu_B = eh/2m_0$ with $m_0$ represents the mass of electron. **S** and **L** are respectively depicting the spin and orbital angular momentum operators.

Next, to evaluate the applicability of k.p model to our current system of $V_2O_3/WSe_2$ vdW heterostructure, we further investigated the SOC-PROCAR results at the band-edges. The orbital-resolved calculations reveal that both the CBM and VBM remain predominantly derived from the W-d orbitals, while the contribution from the V-d orbitals is negligible. However, compared with pristine $WSe_2$, the orbital contribution is partially modified due to the interfacial hybridization and electronic-state reconstruction. Therefore, the orbital characteristics obtained from pristine $WSe_2$ cannot be directly used to assign the exact orbital angular momentum quantum numbers associated with the band-edge states. Consequently, the conventional orbital angular momentum values of pristine $WSe_2$ ($L_z$ = 0 for CBM) and ($L_z$ = 2 for VBM) as shown in Electronic Supplementary Information Figure ESI-7 can only provide an approximate description of the electronic states in $V_2O_3/WSe_2$ vdW heterostructure. Therefore, the estimated effective magnetic field of 38.61 T corresponding to the valley splitting of 10.41 meV should be regarded as a qualitative measure of the magnetic proximity effect rather than a quantitative accurate value. Nevertheless, the sizeable valley splitting directly obtained from the first-principles calculations confirms the presence of strong exchange-induced valley polarization in the $V_2O_3/WSe_2$ vdW heterostructure.

Next, we applied an external electric field in the positive and negative direction and explored the valley splitting. Interestingly, we found more enhancement in the valley splitting in positive direction as compared in the negative direction. The results of valley splitting are given in Table 1 for both positive and negative electric field direction. From Table 1, the negative electric field first weakens the valley splitting and then enhances it due to the charge redistribution and an increase strength of spin-orbit interactions. Therefore, overall, we found a competition between the charge redistribution and interfacial exchange coupling under the external electric field.

**Table 1: Lattice constant (a), interlayer distance ($d_0$), total energy difference (ΔE), band gap energies $E_g$, energy difference at $K^+$ and $K^-$ (Δ), and magnetocrystalline anisotropy energy (MAE) for positive/negative electric field.**

| | Pristine | 0.1 | 0.2 | 0.3 | 0.4 | 0.5 | 0.6 |
|---|---|---|---|---|---|---|---|
| ***Positive*** | | | | | | | |
| **$E_0$ (meV)** | 314.88 | 341.17 | 312.64 | 311.76 | 310.68 | 309.6 | 348.3 |
| **$d_0$ (Å)** | 3.19 | 3.148 | 3.143 | 3.139 | 3.135 | 3.131 | 3.127 |
| **$E_g$ (eV)** | 0.030 | 0.019 | 0.029 | 0.029 | 0.032 | 0.033 | 0.032 |
| **Δ (mev)** | 10.41 | 11.18 | 12.95 | 15.91 | 18.50 | 27.12 | 38.14 |
| **MAE (meV)** | 0.31 | 0.37 | 0.41 | 0.41 | 0.42 | 0.42 | 0.42 |
| ***Negative*** | | | | | | | |
| **$E_0$ (meV)** | 314.88 | 341.17 | 312.64 | 311.76 | 310.68 | 309.6 | 348.3 |
| **$d_0$ (Å)** | 3.19 | 3.156 | 3.199 | 3.201 | 3.161 | 3.244 | 3.275 |
| **$E_g$ (eV)** | 0.030 | 0.031 | 0.029 | 0.030 | 0.031 | 0.032 | 0.030 |
| **Δ (mev)** | 10.41 | 10.06 | 9.51 | 11.83 | 14.23 | 16.84 | 23.45 |
| **MAE (meV)** | 0.31 | 0.39 | 0.38 | 0.37 | 0.36 | 0.36 | 0.34 |

To further understand the interfacial interaction of the most stable configuration S1, we calculate and investigate the charge redistribution at the interface of the $V_2O_3/WSe_2$ heterostructure by using the relation

$$\Delta_\rho = \rho_{Het} - \rho_{WSe_2} - \rho_{V2O3} \ (3)$$

Where $\rho_{Het}$, $\rho_{WSe_2}$, and $\rho_{V2O3}$ are respectively the charge density of the vdW $V_2O_3$/$WSe_2$ vdW heterostructure, $WSe_2$ and $V_2O_3$ monolayers. The calculated charge density difference as denoted by $\Delta_\rho$ reveals a noticeable charge redistribution at the interface of the two monolayers for the electron depletion and accumulation as depicted in Figure 3. The yellow and cyan colors represent the charge accumulation and depletion. To further confirm the transfer of charges from one layer to the other, we also performed the Bader charge analysis.[61,62] The negative and positive values represent the transfer of charge from $V_2O_3$ to $WSe_2$ monolayers or vice versa respectively. A charge of 0.5924 e is lost by the W atoms of the $WSe_2$ monolayer, while the V and O atoms respectively gain a charge of 0.298 and 1.140 e. Further, the Se atoms towards the $V_2O_3$ monolayer also gain a charge of 0.245 e, while the Se atoms on the opposite side have an almost negligible contribution in the charge redistribution as shown in Figure 3. Interestingly, Figure 3(a) also reveals the presence of an intrinsic built-in electric field originating from interfacial charge transfer between the $V_2O_3$ and $WSe_2$ monolayers. This charge redistribution leads to an asymmetric electrostatic potential across the interface, confirming strong interlayer coupling in the heterostructure. Consequently, a magnetic proximity effect emerges at the interface, where the magnetic ordering of the $V_2O_3$ monolayer induces a finite magnetic polarization in the adjacent $WSe_2$ layer. Notably, this induced magnetization is primarily localized on the Se atoms facing the $V_2O_3$ monolayer, while the Se atoms on the opposite side exhibit negligible magnetic moments due to the weak interaction with the ferromagnetic $V_2O_3$ monolayer. Further, this charge redistribution also modifies the electronic and magnetic properties of the $V_2O_3$/$WSe_2$ heterostructure.

Later, we also applied the external perpendicular electric field and explored the effect of electric field over the charge density difference as shown in Figure 3 (b) and (c) respectively for $\pm 0.6$ V/Å. Similarly, we also study the effect of electric filed over the Bader charge analysis as given in Table 2.

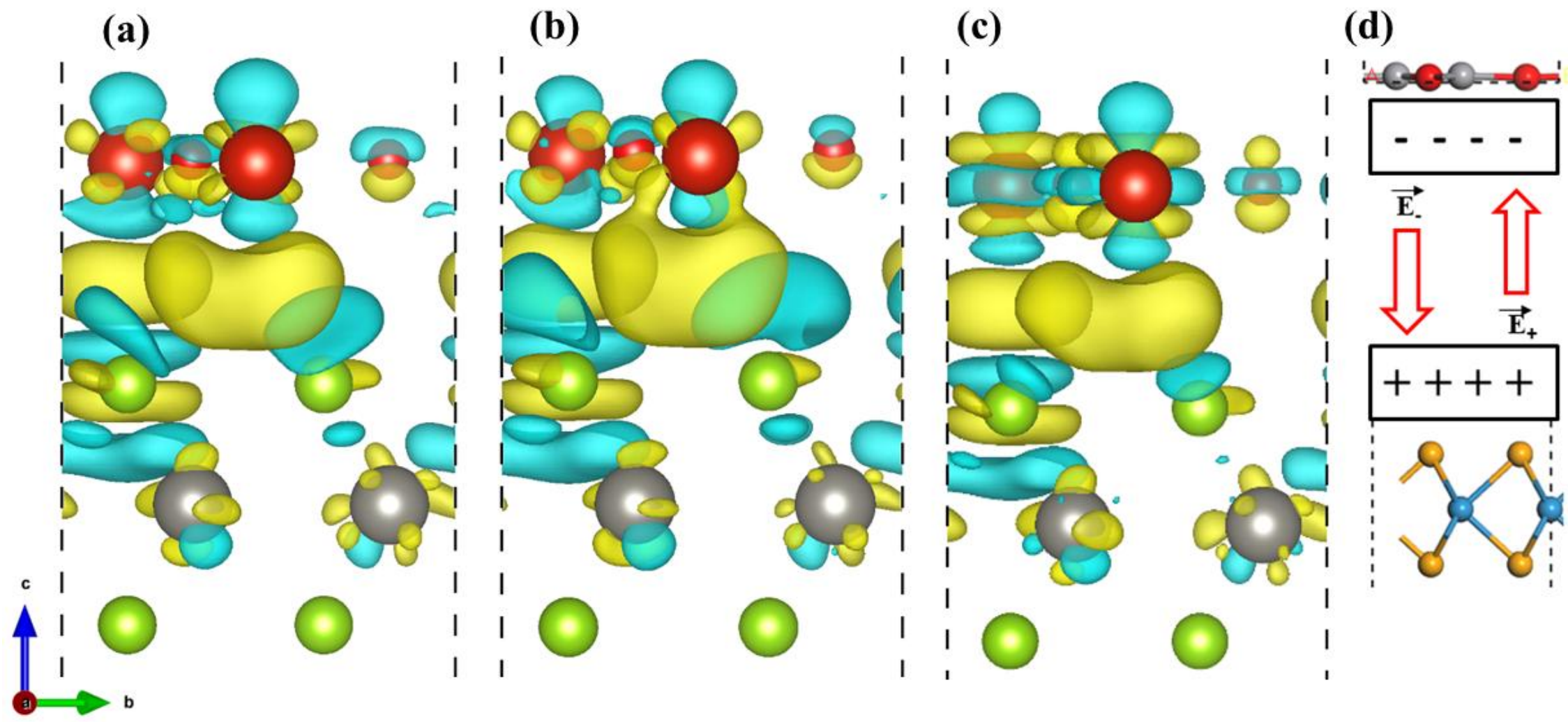

**Figure 3:** Charge density difference of $V_2O_3$/$WSe_2$ vdW heterostructure of **(a)** pristine, **(b)** + 0.6 and **(c)** – 0.6 V/Å electric field. The cyan and yellow color shows the charge depletion and accumulation with an iso-surface value of $\pm 0.0003$ e Å$^{-3}$. **(d)** shows the direction of electric field.

Interestingly, the Bader charge transfer has been confirmed from the charge density difference not only at zero electric field but also at high electric field.

**Table 2: Transfer of charges at different electric fields from $WSe_2$ (Positive) to $V_2O_3$ monolayer and vice versa (Negative). The plus and minus values for charge transfer show the charge accumulation and depletion respectively.**

| | ***Positive*** | | | | ***Negative*** | | | |
|---|---|---|---|---|---|---|---|---|
| | **W** | **Se** | **V** | **O** | **W** | **Se** | **V** | **O** |
| **0.1** | 1.775 | -0.932 | 0.052 | -1.339 | 0.536 | -0.303 | -0.314 | -1.115 |
| **0.2** | 1.768 | -0.923 | 0.049 | -1.344 | 0.686 | -0.276 | -0.327 | -1.105 |
| **0.3** | 1.778 | -0.929 | 0.048 | -1.347 | 0.688 | -0.267 | -0.297 | -1.122 |
| **0.4** | 1.775 | -0.932 | 0.043 | -1.350 | 0.596 | -0.227 | -0.271 | -1.096 |
| **0.5** | 1.781 | -0.941 | 0.042 | -1.351 | 0.671 | -0.298 | -0.312 | -1.105 |
| **0.6** | 1.777 | -0.943 | 0.040 | -1.354 | 0.662 | -0.283 | -0.284 | -1.084 |

### D. **Magnetic Properties**

To study the effect of electric fields, first we need to investigate the magnetic ground state of the pristine system. In the vdW $V_2O_3/WSe_2$ heterostructure, we consider both the ferromagnetic (FM) and the Neel anti-ferromagnetic (AFM) configurations as shown in Figure 1 (a) by the blue arrows. For the FM and AFM spin configuration we only consider a unit cell of the $V_2O_3/WSe_2$ heterostructure. Where V atoms mainly contributed to the total magnetic moment while other atoms have negligible contribution to total the magnetic moment. According to Hund's rule, each electron in V atom occupies the $t_{2g}$ triplet state, which results in a 2 $\mu_B$ magnetic moment. In the case of AFM, one V atom has a magnetic moment of +2 while the other has -2 $\mu_B$. Therefore, the total magnetic moment of $V_2O_3/WSe_2$ vdW heterostructure is zero. Further, in the case of FM, each V atom has a magnetic moment of 2 $\mu_{B.}$ Therefore, the total magnetic moment is 4 $\mu_B$. Further, consistent with the Goodenough-Kanamori-Anderson (GKA) superexchange mechanism,[63–65] the magnetic moments mainly originate from the V atoms ~1.971, while small, induced moments -0.14, 0.16 and 0.011 $\mu_B$ appear on neighboring O, W, and interfacial Se atoms due to hybridization. A 0.005 $\mu_B$, magnetic moment is also induced on Se atoms opposite to the $V_2O_3$ monolayer. The difference of energies between the antiferromagnetic (AFM) and ferromagnetic (FM) configuration states can be evaluated by using the relation

$$E_0 = E_{AFM} - E_{FM} \quad (4)$$

where $E_0$, $E_{AFM}$ and $E_{FM}$ respectively stands for the energy of current, anti-ferromagnetic and ferromagnetic magnetic ground states of the $V_2O_3/WSe_2$ vdW heterostructure. Remarkably, Table 1 also depicts, that an increase in external electric field from/to $WSe_2$ to $V_2O_3$ monolayer does not robustly affect the ferromagnetic ground state of the $V_2O_3/WSe_2$ vdW heterostructure.

Later, we also calculated the magneto-crystalline anisotropy energy (MAE) by calculating the non-collinear energies, including the spin-orbit (SOC) interaction along the in-plan [1 0 0] and out-of-plan [0 0 1] directions which are called in-plan and out of plane directions. As after applying SOC, the investigation of MAE can be conducted from the expression.

$$MAE = \xi^2 \sum_{u,o;\alpha,\beta}(2\delta_{\alpha\beta} - 1)\,[\frac{|<u,\alpha|\hat{L}z|o,\beta>|^2 - |<u,\alpha|\hat{L}x|o,\beta>|^2}{\varepsilon_{u,\alpha} - \varepsilon_{o,\beta}}] \qquad (5)$$

where $\xi$ shows the SOC coupling, while $\varepsilon_{u,\alpha}$ and $\varepsilon_{o,\beta}$ stands for unoccupied and occupied states energy levels both in the spin up (α) and down (β) orientations. Moreover, while the magnetic moment is largely governed only by the spin population difference and is thus insensitive to the orbital character, while the MAE strongly sensitive to the orbital character due to the SOC mechanism. Interestingly, we noted a giant MAE in the pristine $V_2O_3/WSe_2$ heterostructure. Further, the MAE of $V_2O_3/WSe_2$ heterostructure increases with an increase in the positive or electric field while it is decrease in increase of the electric field in the opposite direction as shown in Table 1. We also investigate the orbital resolved contribution to the total MAE and found that in pristine $V_2O_3/WSe_2$ vdW heterostructure only $V_d$ orbital contributes to the total magnetic anisotropy with an average value of 0.282 meV per V atom, while the other atoms contribution is almost negligible as shown in Figure 4 (b). Interestingly, the dominant contribution to the perpendicular MAE originates from the coupling between $d_{yz}$ and $d_z{}^2$, while the next largest contribution arises from the interaction between $d_{yz}$ and $d_{xz}$ orbitals. Furthermore, Figure 4 (a) confirms that, the $d_{xy}$ and $d_{xz}$ orbitals contribute negligibly to the total MAE. Similarly, for the other constituent atoms (O, W, and Se), we examined their orbital-resolved contribution to the total MAE. The results shows that all these three atoms contribute negligibly to the total MAE, with a marginally contribution from W atom. However, in case of Se atom, the ($p_z$, $p_y$) orbital coupling favors the out of plane MAE, whereas the ($p_x$, $p_y$) orbital pair coupling of the O atoms contributes more significantly to the out-of-plane MAE, as shown in Figure 4 (b). It is important to note that we only investigate the atoms with comparatively large contribution to MAE, which occupy the site encircled as blue, black, red, and green dashed circles in Figure 1 (a) respectively for V, O, W and Se atoms in the crystal structure of $V_2O_3/WSe_2$ heterostructure. Furthermore, the external electric field does not affect the orbital resolved contribution to the total MAE other than the atomic sites, as shown in the Electronic Supplementary Information Figure ESI-8.

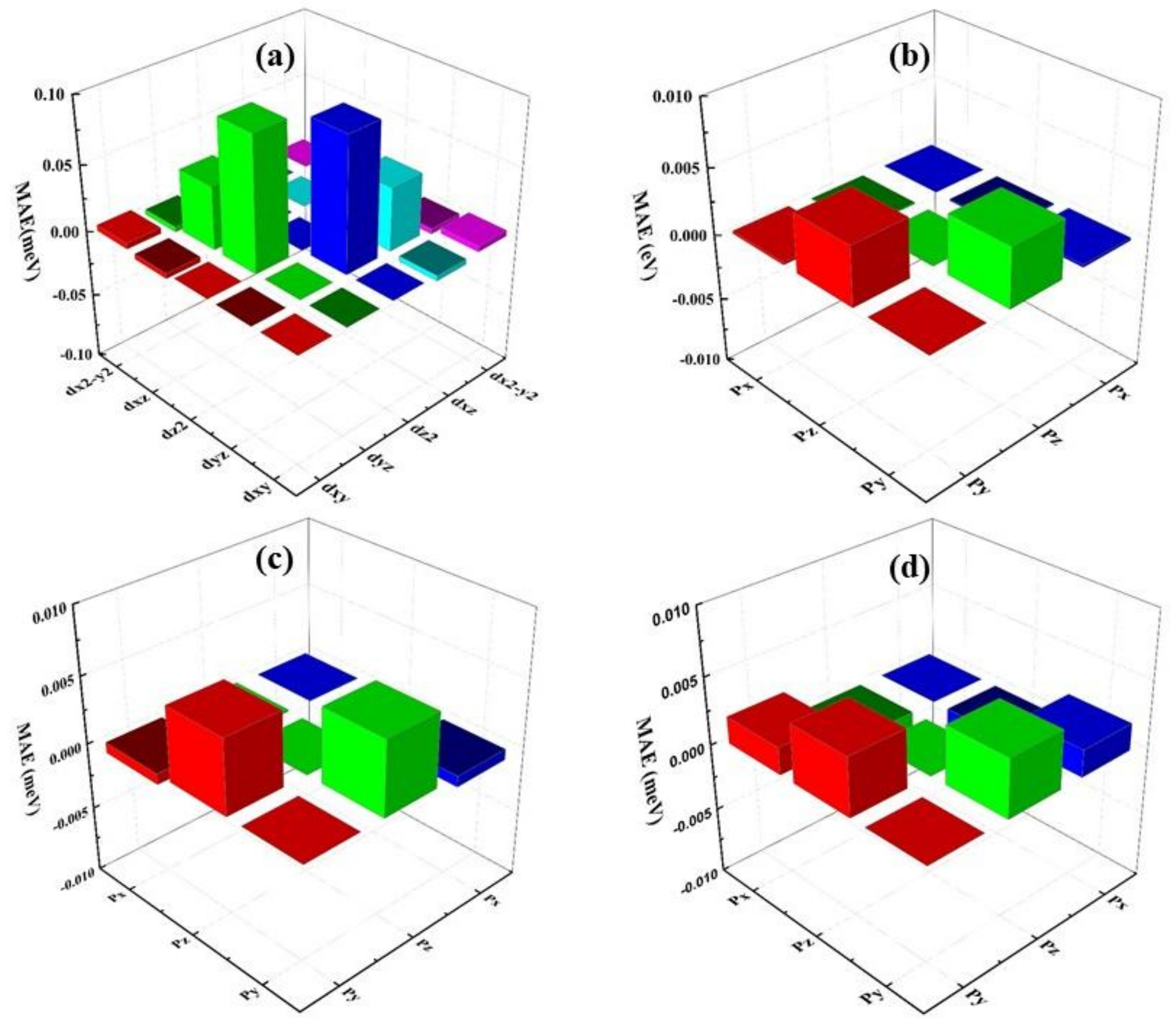

**Figure 4:** SOC resolved MAE in pristine $V_2O_3$/ $WSe_2$ vdW heterostructure of **(a)** V **(b)** O **(c)** W and **(d)** Se respectively.

Due to the electronegativity difference between the two monolayers forming $V_2O_3/WSe_2$ heterostructure, an induce dipole moment is expected at the interface. To understand the origin of the charge transfer, we need to analyze the planar average electrostatic potential and consequently the induced dipole moment as shown in Figure 5. The planer average electrostatic potential is calculated along the z-axis which is in accordance with the in-built or intrinsic electric field as shown in Figure 5 (a). Furthermore, this in-built electric field causes a charge transfer from $V_2O_3$ to $WSe_2$ ML even at zero electric field. Moreover, the planar electrostatic potential also confirms the orientation of both MLs as shown in Figure 5 (a). Besides, both the inversion and time-reversal symmetries are broken by the $V_2O_3$ ferromagnetic ML and thus the valley splitting occurs due to the magnetic proximity effect induced by $V_2O_3$ ML. Interestingly, the in-plane magnetic proximity effect or effective magnetic field has no contribution to the valley splitting, while the perpendicular effective magnetic field has a negligible effect over the valley splitting.

The difference in the electronegativity between the two layers not only results in an induced dipole moment but also a charge density difference. In the current calculations, we calculate the dipole moment along the z-direction, perpendicular to the $V_2O_3/WSe_2$ vdW heterostructure using the following equation,

$$D_M = -\int \rho(z) z dz + \sum N_i e z_i \quad (6)$$

In the above equation, *ρ(z)* represents the valence electron density, *z* is the co-ordinate, *Ni* shows the atomic number of each ion *i*, and *e* is the charge of one proton. The charge redistribution at the interface of the heterostructure resulted in an induced dipole moment of 0.023 eÅ. Interestingly, as the external perpendicular electric field increases either in the positive or negative direction, the dipole moment also increases as shown in Figure 5 (b). This is due to the external perpendicular electric field and the internal in-built electric field, which is mainly due to the difference in the electronegativity between the two layers of the $V_2O_3/WSe_2$ vdW heterostructure. Further, the enhancement in the dipole moment with increasing positive electric field is due to the more inclination of the electronic charge from $WSe_2$ towards $V_2O_3$ monolayer. Similarly, in the opposite direction, the dipole moment is still enhanced but with a negative sign only. A small difference in the dipole moment is only due to built-in electric field, because at zero electric field we found a dipole moment from $WSe_2$ towards $V_2O_3$ monolayer, it means an intrinsic electric field is induced due to the $V_2O_3$ ferromagnetic monolayer.

Similarly, to calculate the work function ($\Phi$), we use the following relation

$$\Phi = E_v - E_F \quad (7)$$

Where $E_v$ is the vacuum energy extracted from the electrostatic potential profile along the *z* direction and $E_F$ is the fermi energy of the $V_2O_3/WSe_2$ vdW heterostructure. Due to the asymmetric in dipole moment in the $V_2O_3/WSe_2$ vdW heterostructure, the work function exhibits a strong increase in the positive electric field direction than in the negative direction. Consequently, a significant potential difference is established across the $V_2O_3/WSe_2$ vdW heterostructure providing further evidence for the existence of a built-in electric field as shown in Figure 5 (c). We also found a significant difference for the vacuum energy ($F_v$) for positive and negative electric fields as depicted in Figure 5 (d). Further, from equation 6, at zero electric field the charges will transfer to a layer with high work function. It is clear from Table 2, that the charge transfer is greater due to the large value of the work function of $WSe_2$ while it goes down due to the negative electric field.

**E. Curie temperature**

Next, we also calculated the Curie Temperature of $V_2O_3/WSe_2$ vdW heterostructure using the Metropolis Monte Carlo (MC) approach among the different available approaches. For this purpose, we use the VAMPIRE simulation package[66]. Here we performed the temperature dependent magnetization curve by using the Metropolis Monte Carlo (MC) simulation technique which is coded in the VAMPIRE software package and then we extracted the Curie temperature ($T_c$). Further, the following classical spin Heisenberg model can be used as written below,

$$H = -\sum_{i,j} J\, m_i . m_j \quad (8)$$

In the above equation, J shows the energy parameter for exchange coupling of the first nearest neighbor atom and $m_i$, $m_j$ depicts the magnetic moments of the nearest neighbors' atoms. Later, the $J_i = \frac{\Delta E}{Nm^2}$ can be used for calculating the exchange coupling energy parameter. Where N shows the number of magnetic atoms in a unit cell and m represents the magnetic moment of each magnetic atom in the $V_2O_3/WSe_2$ vdW heterostructure unit cell.

Now, to run the Monti-Carlo simulation code, we need a large enough supercell of 50 x 50 x 1 so that to minimize the finite-size effects and obtaining a long-range magnetic correlations and accurate thermodynamic properties like Curie temperature. Further, the magnetization curve can be fitted by using the Curie-Bloch relation as given below.

$$m(T) = \left[1 - \frac{T}{T_C}\right]^{\beta} \quad (9)$$

In the above equation, $\beta$ is the critical exponent having different values is used for fitting the temperature dependent magnetization curve as shown in Figure 6. Here the value for this critical exponent is 0.49.

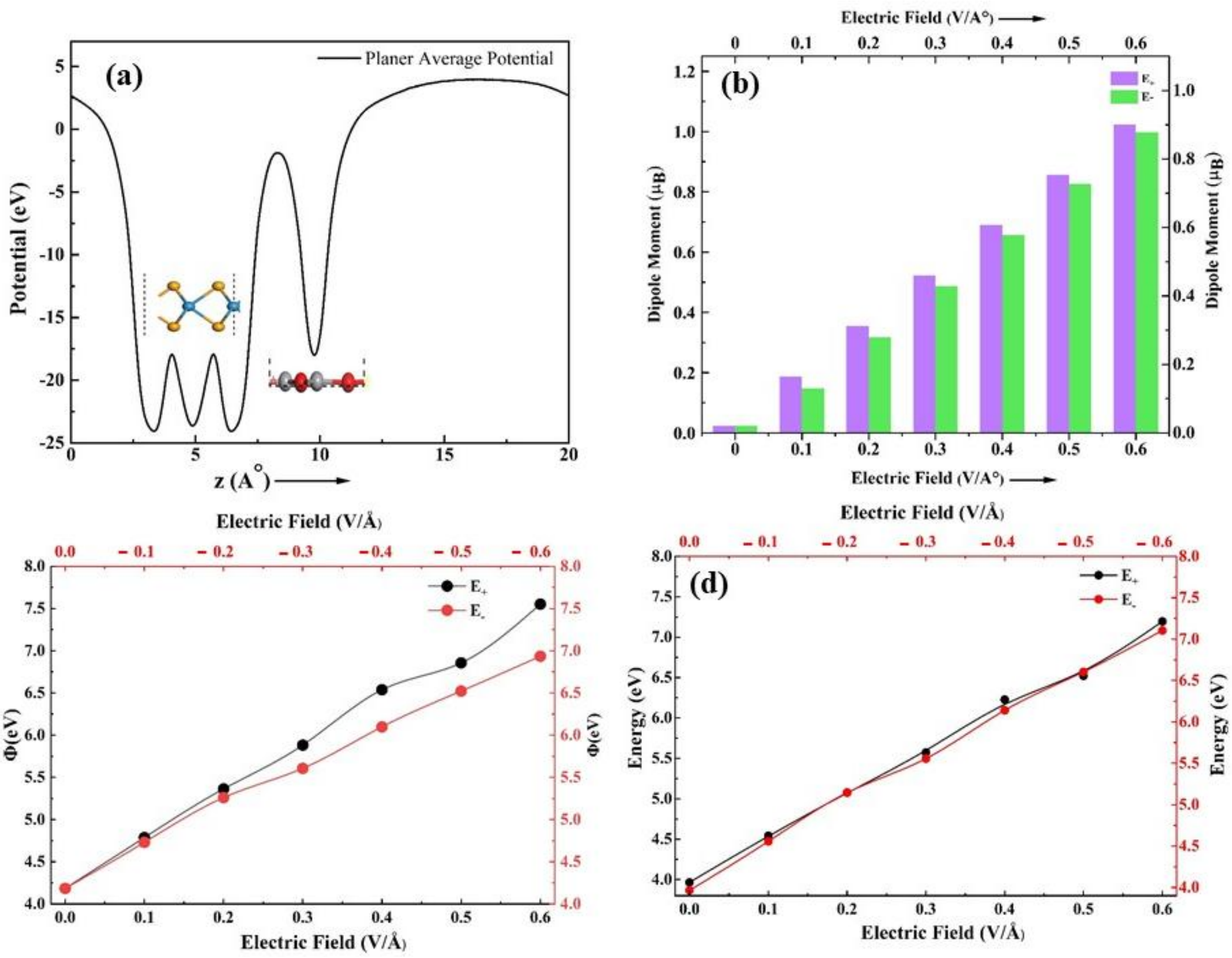

**Figure 5: (a)** Electrostatic potential for pristine $V_2O_3/WSe_2$ vdW heterostructure, Electric field dependent **(b)** dipole moment, **(c)** work function (**Φ**) and **(d)** vacuum energy (**$F_v$**) of $V_2O_3/WSe_2$ vdW heterostructure.

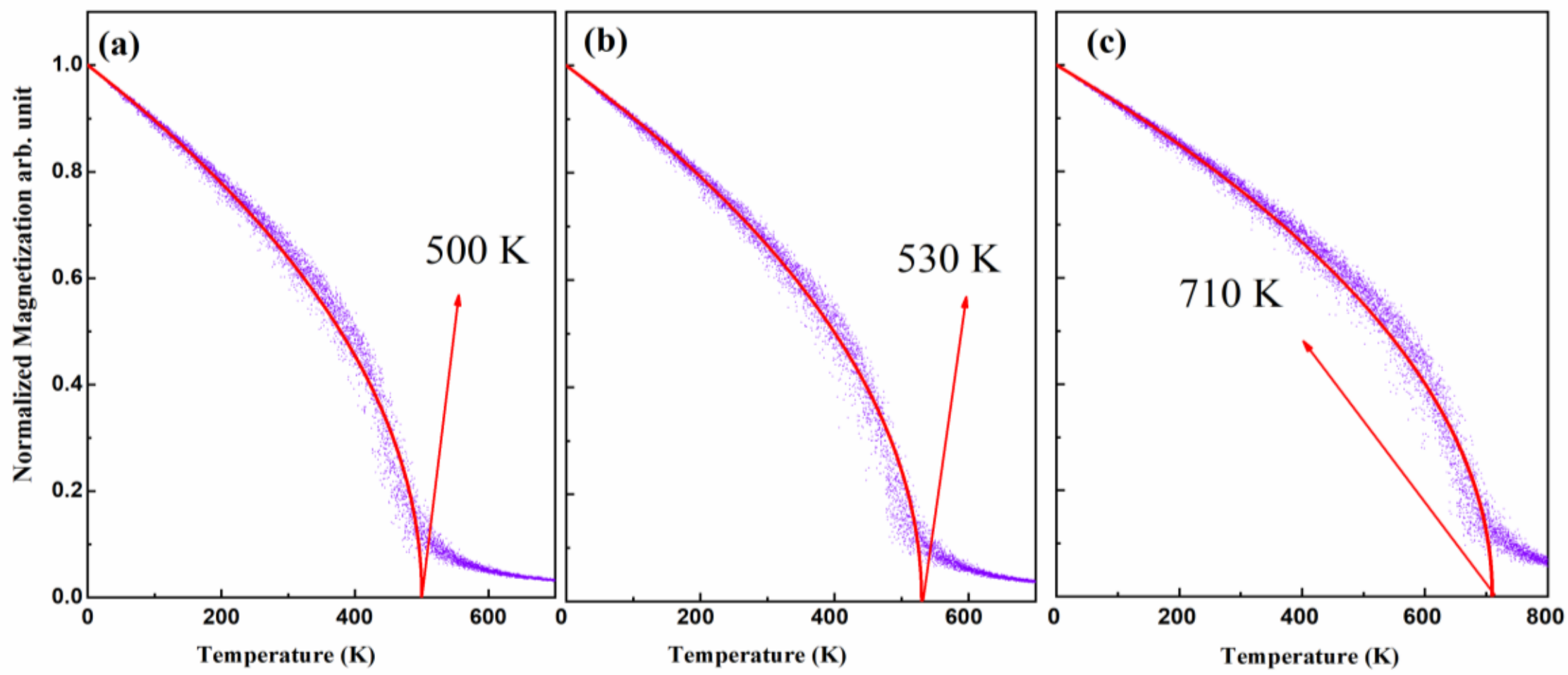

**Figure 6:** Curie temperature of the $V_2O_3/WSe_2$ heterostructure, respectively at pristine (a) + 0.6 V/Å (b) - 0 .6 V/Å. The purple dotted curve shows the temperature dependent magnetization curve, while the red line shows the fitted result with Eq. (9).

This Figure shows that initially, the magnetization curve retains at high values due to the high spin state in the low temperature range due to its stable ferromagnetic ordering and gradually decreases with increase in temperature, eventually it drops zero near the critical temperature. From this Monte Carlo simulations, we obtained a Curie temperature of approximately 500 K for the pristine $V_2O_3/WSe_2$ vdW heterostructure. To verify the implementation of the Heisenberg Monte Carlo methodology, we also estimated the Curie temperature of $CrI_3$ monolayer and found it 46 K, which agree well with the experimentally reported value of 45 K.[67] It is worth mentioning that our intention of this comparison is solely as a methodological benchmark of the Monte Carlo implementation and should not be regarded as a universal validation of the Monti Carlo simulation approach for all the ferromagnetic oxide materials. Further, this comparison shows a good agreement between the experimental and theoretical Curie temperature of $CrI_3$ shows that the current adopted simulation framework can reasonably capture the magnetic phase transition behavior in well-established two-dimensional magnetic materials.

Furthermore, we also compared our results with the Mean Field Theory (MFT) approach and found that the MFT results are overestimated. Later, we also calculated the Curie temperature at high external electric field as shown in Figure 6 (b)-(c) for 0.6 V/Å both in the positive and negative directions. Interestingly, we found that Curie temperatures at +0.6 and -0.6 V/Å are respectively 530 and 710 K. We also calculate the Curie temperature at all the remaining steps as given in the Electronic Supplementary Information Table ESI-1. Similarly, the information about the magnetic moment and exchange energy are also listed in Table ESI-1.

## IV. Discussion

In the current study, besides the structural, electronic, and magnetic properties, we also explore the valleytronics properties of $V_2O_3$/$WSe_2$ vdW heterostructure. The $V_2O_3$/$WSe_2$ vdW heterostructure had a direct band gap of 0.02 eV. Applying spin-orbit interactions, the magnetic proximity effect causes a giant valley splitting of 10.41 meV in the $V_2O_3$/$WSe_2$ vdW heterostructure. Similarly, the non-collinear total energy calculations confirm an out of plane magnetocrystalline anisotropy with a prominent value of 0.31 meV. Further, using Metropolis Monte Carlo (MC) simulations, we found a Curie temperature of 500 K in the pristine $V_2O_3$/$WSe_2$ vdW heterostructure. We apply an external electric field in the perpendicular direction to the $V_2O_3$/$WSe_2$ vdW heterostructure and explore the effect of electric field over the structural, electronic, and magnetic properties of the $V_2O_3$/$WSe_2$ vdW heterostructure. Interestingly, we found that ferromagnetism remains preserved irrespective of the electric field direction. Further, an abrupt enhancement in valley splitting of the $V_2O_3$/$WSe_2$ vdW heterostructure has also been found. The badder charge analysis confirmed that the charge transfers from/to $V_2O_3$ to/from $WSe_2$ monolayers. This charge transfer induces a in-built electric field at the interface, which can either enhances or weaken the effect of the external applied electric field in the parallel/opposite directions. Consequently, the superexchange interaction described by the Goodenough-Kanamori-Anderson rule is further strengthened, leading to the induction or modification of the magnetic moments of the constituent atoms in the $V_2O_3$/$WSe_2$ vdW heterostructure. This phenomenon is commonly referred to as magnetic proximity effect. Therefore, the applied external electric field plays a decisive role in controlling the electronic, magnetic, and structural properties of $V_2O_3$/$WSe_2$ vdW heterostructure. Therefore, our studies suggest that the external electric field is a prominent technique to control the spin and valleytronics applications of an oxide-based vdW heterostructure.

## AUTHOR INFORMATION

### Author Contributions:

J.J and G.Q supervised the project. F.S conceived the idea, did the DFT calculations, and plotted the results. All the authors carefully check the results and help with writing and editing the manuscript.

### Acknowledgments:

This work is supported by the Industrial Technology Innovation Program (RS-2024-00435432: Development of hydrogen production technology through ammonia decomposition using a plasmonics-based light system with high performance and long-term durability), funded by the Ministry of Trade Industry & Energy (MOTIE, Korea). This research was supported by Brain Pool program funded by the Ministry of Science and ICT through the National Research Foundation of Korea (RS-2025-25396197). G.Q. is supported by the National Key R&D Program of China (2023YFB2408100), the Fundamental Research Funds for the Central Universities (Grant Nos. 531119200237), the Guangdong Basic and Applied Basic Research Foundation (2025A1515012237), the State Key Laboratory of Robotics and Systems at Harbin Institute of Technology (SKLRS-2025-KF-10), the National Major Science and Technology Project of China (Grant No. 2025ZD0717301), and the Open Project of Shaanxi Key Laboratory of Gear Transmission (Grant No. SKLGT-2024-005).

### Corresponding Authors

*E-mail: jkjang@pusan.ac.kr (J. J), gzqin@hnu.edu.cn (G. Q);

### Conflict of interest

The authors have no conflict of interest.